\documentclass
[aps,prl,amsmath,amsfonts,amssymb,twocolumn,superscriptaddress,showpacs]{revtex4}%
\usepackage{times}
\usepackage{amsfonts}
\usepackage{amsmath}
\usepackage{amssymb}
\usepackage{graphicx}%
\providecommand{\U}[1]{\protect\rule{.1in}{.1in}}

\begin{document}

\title{Exact dynamics of single qubit gate fidelities under \\ the measurement-based quantum computation scheme}

\author{L. G. E. Arruda}
 \email{lgarruda@ursa.ifsc.usp.br}
\affiliation{Instituto de F\'{\i}sica de S\~{a}o Carlos, Universidade de S\~{a}o Paulo, P.O. Box 369, 13560-970, S\~{a}o Carlos, SP, Brazil}

\author{F. F. Fanchini}
\email{fanchini@fc.unesp.br.}
\affiliation{Departamento de F\'{\i}sica, Faculdade de Ci\^{e}ncias, UNESP, CEP 17033-360, Bauru, SP, Brazil}

\author{R. d. J. Napolitano}
\affiliation{Instituto de F\'{\i}sica de S\~{a}o Carlos, Universidade de S\~{a}o Paulo, P.O. Box 369, 13560-970, S\~{a}o Carlos, SP, Brazil}

\author{J. E. M. Hornos}
\affiliation{Instituto de F\'{\i}sica de S\~{a}o Carlos, Universidade de S\~{a}o Paulo, P.O. Box 369, 13560-970, S\~{a}o Carlos, SP, Brazil}
\date{\today}

\author{A. O. Caldeira}
\affiliation{Instituto de F\'{\i}sica Gleb Wataghin, Universidade Estadual de Campinas, P.O. Box 6165, CEP 13083-970, Campinas, SP, Brazil}

\begin{abstract}
Measurement-based quantum computation is an efficient model to perform universal computation. Nevertheless, theoretical questions have been raised, mainly with respect to realistic noise conditions. In order to shed some light on this issue, we evaluate the exact dynamics of some single qubit gate fidelities using the measurement-based quantum computation scheme when the qubits which are used as resource interact with a common dephasing environment. We report a necessary condition for the fidelity dynamics of a general pure $N$-qubit state, interacting with this type of error channel, to present an oscillatory behavior and we show that for the initial canonical cluster state the fidelity oscillates as a function of time. This state fidelity oscillatory behavior brings significant variations to the values of the computational results of a generic gate acting on that state depending on the instants we choose to apply our set of projective measurements. As we shall see, considering some specific gates that are frequently found in the literature, neither fast application of the set of projective measurements necessarily implies high gate fidelity, nor slow application thereof necessarily implies low gate fidelity. Our condition for the occurrence of the fidelity oscillatory behavior shows that the oscillation presented by the cluster state is due exclusively to its initial geometry. Other states that can be used as resources for measurement-based quantum computation can present the same initial geometrical condition. Therefore, it is very important for the present scheme to know when the fidelity of a particular resource state will oscillate in time and, if this is the case, what are the best times to perform the measurements.
\end{abstract}


\maketitle

\section{INTRODUCTION}

Quantum computation and information theory are research areas which have developed amazingly fast. This rapid and accelerated development is due to the promises of qualitatively new modes of computation and communication based on quantum technologies, that in some cases are much more powerful than their classical counterparts, and also due to the impact that these new technologies can provide to society. Advances occur in different directions and the theoretical proposals as well as the experimental achievements involves fundamental concepts that live at the heart of quantum mechanics. One of the most remarkable examples is the measurement-based quantum computation \cite{briegel01, briegel02, briegel03, newoneway, newoneway2}. While the standard quantum computation is based on sequences of unitary quantum logic gates, the measurement-based quantum computation is realized using only local conditioned projective measurements applied to a highly entangled state called the cluster state. As it is entirely based on local measurements instead of unitary evolution, the computation is inherently irreversible in time. Recently, this one-way quantum computation has attracted a lot of attention of the scientific community \cite{teoricos} and has been studied considering: (i) various important aspects which might influence it, such as decoherence during the computation \cite{decoOW} and (ii) novel schemes of implementation \cite{newoneway, novel}. In addition, since the technical requirements for the latter can be much simpler than those for the standard circuit model, it has been realized in several experiments \cite{experimentais}.

The noise process, that emerges from the inevitable interaction of the qubits with their environment, remains one of the major problems to be overcome before we shall be able to manufacture a functional quantum computer. The interaction of the qubits with the environment, which depends on the quantum computer architecture, is a crucial problem that is far from being well understood. However, it is known that this interaction can result in non-monotonical dynamics of the density matrix coherences \cite{quiroga, breuer,bellomo} and a complete understanding of this behavior is a matter of great importance. Moreover, this non-monotonical behavior, is a subject of broad interest \cite{wolf, breuer01, laine, breuer02}, since this peculiar property brings up unexpected dynamics for quantum fidelity \cite{rajagopal} as well as for  quantum entanglement \cite{bellomo, he, xiao, rivas} and quantum discord \cite{fanchini}. When an open quantum system interacts with the outside world there are two main effects that have to be considered: the relaxation and the decoherence processes. The relaxation process is associated with an expected loss of energy of the initial state of the system which happens at the rate $\tau^{-1}_{rel}$, where the time scale $\tau_{rel}$ is known as the relaxation time scale of the system. On the other hand, the decoherence process is associated with the reduction of purity of this physical state and takes place within a time scale  $\tau_{dec}$. Depending on the physical system one considers, the time scale $\tau_{dec}$ can be much shorter than $\tau_{rel}$, making quantum computers more sensitive to decoherence than to relaxation process. This is exactly the situation we are going to address in this paper.

Based on the foregoing reasoning, we use an exact solvable model \cite{quiroga} to calculate the dissipative fidelity dynamics in a linear measurement-based quantum computer (MBQC) \cite{briegel01} composed by few qubits which interact collectively with a dephasing environment. Although this kind of model does not describe relaxation processes, it does indeed adequately describe decoherence effects. The interaction Hamiltonian is given by $\sigma^{\left(T\right)}_{z}\otimes B$, where $\sigma^{\left(T\right)}_{z}\equiv\sum^{N}_{n = 1}\sigma^{\left(n\right)}_{z}$ is the total azimuthal angular momentum in a system of $N$ qubits, and $B=\hbar\sum_{k}\left(g_{k}a^{\dag}_{k} + g^{*}_{k}a_{k}\right)$ is the operator that acts only on the environmental degrees of freedom. We show that for any initial state given by an eigenstate superposition of the $\sigma^{\left(T\right)}_{z}$ operator, whose eigenvalues are different in modulus, the fidelity exhibits a non-monotonical character. To be more precise, suppose we have a system whose initial state can be written in a suitable basis as $\vert\psi\rangle = \sum^{1}_{m_{1}=0}\sum^{1}_{m_{2}=0}\dots\sum^{1}_{m_{N}=0}c_{m_{1},m_{2},\dots ,m_{N}}\vert m_{1}, m_{2},\dots , m_{N}\rangle$ where $\sigma^{\left(T\right)}_{z}\vert m_{1}, m_{2},\dots , m_{N}\rangle = M\vert m_{1}, m_{2},\dots , m_{N}\rangle$, with $M = m_{1} + m_{2} + \dots + m_{N}$; if the vectors that characterize the state $\vert\psi\rangle$ are such that their $\sigma^{\left(T\right)}_{z}$ eigenvalues are all equal in modulus, i.e., if their $\vert M\vert$'s are all equal, the fidelity dynamics does not exhibit a non-monotonical shape, but if at least one vector that composes the state $\vert\psi\rangle$ has a $\vert M\vert$ different from the other vectors we can observe a non-monotonical behavior of the fidelity dynamics. This condition reveals itself to be necessary for the fidelity to present an oscillatory shape, and,  remarkably, it does not depend on the initial entanglement (depending only on the initial configuration of the state of the system).
Furthermore, we study the implications of this oscillatory behavior to the MBQC where a sequence of local projective measurements are applied on the qubits to implement a quantum gate. As we shall see, we can take advantage of the revival times of the coherence of the cluster state to apply the projective measurements at times such that we get the best gate fidelity values. Therefore, we can choose the best possible instants that produce the best possible results. In this sense, fast measurements can result worse than slow, but conveniently applied measurements. That is, under the action of a common dephasing environment, the MBQC can provide better computational fidelities even for slow measurement sequences. This result, as we will see bellow, is a natural consequence of the oscillatory behavior of the density-matrix coherences. To illustrate our finding we examine the fidelity of some single qubit quantum gates that are frequently found in the literature \cite{chuang} developed using the MBQC scheme.

This manuscript is organized as follows. In section II we describe the exact dissipative dynamics of the $N$-qubit system interacting with a common dephasing environment \cite{quiroga}. In section III we show the necessary condition for the non-monotonical behavior of the fidelity to take place and, in section IV, we present the implications of our results to the MBQC. We conclude in section V.

\section{EXACT DISSIPATIVE DYNAMICS}

Consider the following Hamiltonian of a system composed of $N$ qubits interacting with a common dephasing environment:
\begin{equation}
H = \sum^{N}_{n=1}\epsilon_{n}\sigma^{\left(n\right)}_{z} + \sum_{k}\epsilon_{k}a^{\dag}_{k}a_{k} + \hbar\sum_{n,k}\sigma^{\left(n\right)}_{z}\left(g_{k}a^{\dag}_{k} + g^{*}_{k}a_{k}\right),\label{EDD.1}
\end{equation}
where the first two terms account for the free evolution of the qubits and the environment, and the third term describes the interaction between them. The environment operators, $a^{\dag}_{k}$ and $a_{k}$, are the customary creation and annihilation operators which follow the Heisenberg's algebra $\left[a_{k},a^{\dag}_{k'}\right] = \delta_{k,k'}$. The qubit operator $\sigma^{\left(n\right)}_{z}$ is the Pauli $\sigma_{z}$ operator of the $n$-th qubit. The complex constant $g_{k}$ has dimension of frequency and indicates the coupling between the qubits and the field modes. In addition, $\epsilon_{n} = \hbar\omega^{\left(n\right)}_{0}$ is the difference of energy between the ground and excited levels of the $n$-th qubit and $\epsilon_{k} = \hbar\omega_{k}$ is the energy associated with the $k$-th mode of the field, with $\omega^{\left(n\right)}_{0}$ and $\omega_{k}$ being, respectively, the transition frequency and the field frequency of the $k$-th mode. We assume that initially the state of the qubits and the state of the environment are decoupled, i.e., the total initial state of the system and the environment (which we call the combined system from now on) can be represented by a density operator given by the following tensor product $\rho\left(0\right) = \rho^{Q}\left(0\right)\otimes\rho^{E}\left(0\right)$, where $\rho^{Q}\left(0\right)$ and $\rho^{E}\left(0\right)$ stand for the initial state of the qubits and the initial state of the environment, respectively. We also assume that the environment is initially in thermal equilibrium:
\begin{equation}
\rho^{E}\left(0\right) = \frac{1}{Z}\exp\left(-\beta H_{E}\right), \label{EDD.2}
\end{equation}
where $H_{E} = \sum_{k}\epsilon_{k}a^{\dag}_{k}a_{k}$ is the environment Hamiltonian, $Z = Tr\left[\exp\left(-\beta H_{E}\right)\right]$ is the partition function, and $\beta = 1/k_{B}T$, with $k_{B}$ representing the Boltzmann constant and $T$ being the environment temperature.

Now, let us rewrite the Hamiltonian of Eq. (\ref{EDD.1}) as a sum of two terms, $H = H_{0} + H_{I}$, where $H_{0} = \sum^{N}_{n=1}\epsilon_{n}\sigma^{\left(n\right)}_{z} + \sum_{k}\epsilon_{k}a^{\dag}_{k}a_{k}$ and $H_{I} = \hbar\sum_{n,k}\sigma^{\left(n\right)}_{z}\left(g_{k}a^{\dag}_{k} + g^{*}_{k}a_{k}\right)$. In the interaction picture, $\tilde{H}_{I} = U^{\dag}_{0}H_{I}U_{0}$, that is,
\begin{equation}
\tilde{H}_{I}\left(t\right) = \hbar\sum_{n,k}\sigma^{\left(n\right)}_{z}\left(g_{k}e^{i\omega_{k}t}a^{\dag}_{k} + g^{*}_{k}e^{-i\omega_{k}t}a_{k}\right), \label{EDD.3}
\end{equation}
where $U_{0} = \exp\left(-\frac{i}{\hbar}H_{0}t\right)$.
Moreover, the time evolution operator is given by
\begin{equation}
U_{I}\left(t\right) = \hat{T}\exp\left(-\frac{i}{\hbar}\int^{t}_{0}{\tilde{H}_{I}\left(t'\right)dt'}\right), \label{EDD.4}
\end{equation}
where $\hat{T}$ is the Dyson time-ordering operator. If we substitute Eq. (\ref{EDD.3}) into Eq. (\ref{EDD.4}), we obtain, after some algebra, the following expression for the time evolution operator \cite{quiroga}:
\begin{eqnarray}
U_{I}\left(t\right) &=& \exp{\left\{\sum_{n,k}\left[g_{k}\sigma^{\left(n\right)}_{z}\varphi_{\omega_{k}}\left(t\right)a^{\dag}_{k}-g^{*}_{k}\sigma^{\left(n\right)}_{z}\varphi^{*}_{\omega_{k}}\left(t\right)a_{k}\right]\right\}}\nonumber \\
& \times & \exp\left\{i\sum_{k}\sum_{m,n} \vert g_{k}\vert^{2}\sigma^{\left(m\right)}_{z}\sigma^{\left(n\right)}_{z}s\left(\omega_{k},t\right)\right\}, \label{EDD.5}
\end{eqnarray}
where $\varphi_{\omega_{k}}\left(t\right) = \frac{1-e^{i\omega_{k}t}}{\omega_{k}}$ and $s\left(\omega_{k},t\right) = \frac{\omega_{k}t - \sin\left(\omega_{k}t\right)}{\omega_{k}^{2}}$.

The dynamics of the $N$-qubit system can be written in terms of the density operator of the combined system as follows
\begin{equation}
\rho^{Q}\left(t\right) = {Tr}_{E}\left[U_{I}\left(t\right)\rho^{Q}\left(0\right)\otimes\rho^{E}\left(0\right)U^{\dag}_{I}\left(t\right)\right]. \label{EDD.6}
\end{equation}
Hence, the matrix elements of the reduced density matrix can be expressed as
\begin{equation}
\rho^{Q}_{\left\{i_{n},j_{n}\right\}}\left(t\right) = \langle i_{1}, i_{2},\dots ,i_{N}\vert\rho^{Q}\left(t\right)\vert j_{1}, j_{2},\dots ,j_{N}\rangle, \label{EDD.7}
\end{equation}
where $\left\{i_{n},j_{n}\right\}\equiv\left\{i_{1},j_{1};i_{2},j_{2};\dots;i_{N},j_{N}\right\}$ refers to the $N$-qubit state. Here we have $i_{n},j_{n} = \pm1$, i.e., the $j_{n}$ are the eigenvalues of the $\sigma^{\left(n\right)}_{z}$ Pauli operator associated with the two-level qubit states $\vert0\rangle_{n}$ and $\vert1\rangle_{n}$, and $i_{n}$ are the eigenvalues of the $\sigma^{\left(n\right)}_{z}$ Pauli operator associated with the respective dual space $_{n}\langle0\vert$ and  $_{n}\langle1\vert$.
After some manipulations, we finally obtain the explicit dynamics of the elements of the density matrix of the $N$-qubit system \cite{quiroga}:
\begin{eqnarray}
& &\rho^{Q}_{\left\{i_{n},j_{n}\right\}}\left(t\right) = \exp\left\{-\gamma\left(t,T\right)\left[\sum^{N}_{n=1}\left(i_{n} - j_{n}\right)\right]^{2}\right\} \nonumber \\
&\times &\exp\left\{i\theta\left(t\right)\left[\left(\sum^{N}_{n=1}i_{n}\right)^{2} - \left(\sum^{N}_{n=1}j_{n}\right)^{2}\right]\right\}\rho^{Q}_{\left\{i_{n},j_{n}\right\}}\left(0\right), \nonumber \\
& & \label{EDD.12}
\end{eqnarray}
where
\begin{equation}
\gamma\left(t,T\right) = \sum_{k}\vert g_{k}\vert^{2}c\left(\omega_{k},t\right)\times\coth\left(\frac{\hbar\omega_{k}}{2k_{B}T}\right), \label{EDD.13}
\end{equation}
and
\begin{equation}
\theta\left(t\right) = \sum_{k}\vert g_{k}\vert^{2}s\left(\omega_{k},t\right), \label{EDD.14}
\end{equation}
with $c\left(\omega_{k},t\right) = \frac{1 - \cos\left(\omega_{k}t\right)}{\omega_{k}^{2}}$. In the continuum limit, Eq. (\ref{EDD.12}) reads
\begin{eqnarray}
& &\rho^{Q}_{\left\{i_{n},j_{n}\right\}}\left(t\right) = \exp\left\{-\Gamma\left(t,T\right)\left[\sum^{N}_{n=1}\left(i_{n} - j_{n}\right)\right]^{2}\right\} \nonumber \\
&\times &\exp\left\{i\Theta\left(t\right)\left[\left(\sum^{N}_{n=1}i_{n}\right)^{2} - \left(\sum^{N}_{n=1}j_{n}\right)^{2}\right]\right\}\rho^{Q}_{\left\{i_{n},j_{n}\right\}}\left(0\right), \nonumber \\
& & \label{EDD.15}
\end{eqnarray}
where
\begin{equation}
\Gamma\left(t,T\right) = \int d\omega J\left(\omega\right)c\left(\omega,t\right)\coth\left(\frac{\hbar\omega}{2k_{B}T}\right), \label{EDD.16}
\end{equation}
\begin{equation}
\Theta\left(t\right) = \int d\omega J\left(\omega\right)s\left(\omega,t\right), \label{EDD.17}
\end{equation}
and $J\left(\omega\right) \equiv \sum_{k}\vert g_{k}\vert^{2}\delta\left(\omega - \omega_{k}\right) \equiv \left(dk/d\omega\right)G\left(\omega\right)\vert g\left(\omega\right)\vert^{2}$ is the environment spectral density. This function has a cutoff frequency $\omega_{c}$, whose value depends on the environment and $J\left(\omega\right)\mapsto0$ for $\omega\gg\omega_{c}$. Here we should stress that since we are going to analyze average values involving the reduced density operator - the fidelity in our particular case - we can safely use this reduced state of the system in the interaction picture. In the Schr\"{o}dinger picture there are additional terms  oscillating with frequency $\epsilon_{n}/ \hbar$ in the off-diagonal elements of that operator.

In our model we assume an ohmic spectral density,
\begin{equation}
J\left(\omega\right) = \eta\omega e^{-\omega/\omega_{c}}, \label{EDD.17A}
\end{equation}
where $\eta$ is a dimensionless proportionality constant that characterizes the coupling strength between the system and the environment. Substituting Eq. (\ref{EDD.17A}) into Eq. (\ref{EDD.16}) and Eq. (\ref{EDD.17}) we obtain
\begin{equation}
\Gamma\left(t,T\right) = \eta\int d\omega e^{-\omega/{\omega_c}} \frac{1-\cos\left(\omega t \right)}{\omega}\coth\left(\frac{\hbar \omega}{2 k_{B} T}\right) \label{EDD.17B}
\end{equation}
and
\begin{eqnarray}
\Theta\left(t\right) &=& \eta\int d\omega e^{-\omega/{\omega_c}}\frac{\omega t-\sin\left(\omega t \right)}{\omega}\nonumber\\
&=& \eta\omega_{c} t - \eta\arctan\left(\omega_{c}t\right).\label{EDD.17C}
\end{eqnarray}
The result of the integration in Eq. (\ref{EDD.17B}) is also well-known \cite{legget} and reads:
\begin{equation}
\Gamma\left(t,T\right) = \eta \ln (1+\omega_{c}^{2} t^{2} ) + \eta \ln \left( \frac{\beta \hbar}{\pi t} \sinh\frac{\pi t}{\beta \hbar} \right) \label{EDD.17D}
\end{equation}
where $\beta\equiv 1/k_{B} T$.

It is easy to note, from Eqs. (\ref{EDD.15}) and (\ref{EDD.17D}), that the decoherence effects arising from thermal noise can be separated from those due to the vacuum fluctuations. This separation allows for an exam of different time scales present in the dynamics. The shortest time scale is determined by the cutoff frequency (see Eq. (\ref{EDD.17C}) above ), $\tau_{c}\sim \omega^{-1}_{c}$. The other natural time scale, $\tau_{T}\sim\omega^{-1}_{T}$, is determined by the thermal frequency $\omega_{T} = \frac{\pi k_{B}T}{\hbar}$ (see Eq. (\ref{EDD.17D}) above) which is related to the relaxation of the off-diagonal elements of the density operator. With these two time scales we can define two different regimes of time \cite{quiroga}: the \textit{thermal} regime and the \textit{quantum} regime.

Thermal effects will affect the $N$-qubit system predominantly only for times $t>\tau_{T}$ whereas the quantum regime dominates over any time interval $t$ such that $\tau_{c}<t<\tau_{T}$, when the quantum vacuum fluctuations contribute predominantly. Besides, we can see from Eqs. (\ref{EDD.17C}) and (\ref{EDD.17D}) that for a sufficiently high-temperature environment, i.e., $\hbar\omega_{c}\gtrsim k_{B}T$, the phase damping factor, which is the main agent responsible for the decoherence, behaves as $\Gamma\left(t,T\right)\approx \eta\omega_{T}t$ causing an exponential decay of the off-diagonal elements of the density operator. Moreover, as the  phase factor $\Theta\left(t\right)$ implies an oscillation with frequency $\eta \omega_{c}$, this time evolution is always slightly underdamped. Notice that one should never reach  the overdamped regime since as $\tau_{c}$ is the shortest time scale in the problem it does not make any sense to make $\omega_{T} > \omega_{c} $.

In the low temperature limit, when $ \omega_{c} \gg \omega_{T}$, the relaxation factor behaves as $\Gamma\left(t,T\right)\approx 2 \eta\ln\left(\omega_{c}t\right)$ which leads the off-diagonal matrix elements to an algebraic decay of the form $1/(\omega_{c} t)^{2 \eta}$. In this case, what we have called above the thermal regime is only reached for very long times, $t\gg \omega_{c}^{-1}$.

As we will show below, the quantum regime implies a very different dynamics of the fidelity and shows up peculiar results to the MBQC, where delayed measurements can result in better computation fidelities.

\section{OSCILLATORY FIDELITY DYNAMICS}

In this section we are interested to know when the fidelity dynamics of an $N$-qubit system, interacting collectively with a dephasing environment, will oscillate in time. We introduce, for the quantum regime, a necessary condition for the non-monotonical behavior to be present.
Here, since we always consider a pure state as our initial condition, the fidelity as a function of time, in the interaction picture, is given by
\begin{equation}
F(t)={\rm Tr}\left\{\rho^{Q}\left(0\right)\rho^{Q}\left(t\right)\right\}. \label{FD.1}
\end{equation}

The dynamics of qubits interacting with a common environment is strongly dependent upon the initial condition, and we will show that, for the quantum regime, the fidelity will always present a non-monotonical behavior when the $N$-qubit system is written as a coherent superposition of $\sigma^{\left(T\right)}_{z}$ eigenstates whose eigenvalues are different in modulus.
As we can see from Eq. (\ref{EDD.15}), the second exponential term is responsible for the oscillations and it is identical to the unity when $\left(\sum^{N}_{n=1}i_{n}\right)^{2} = \left(\sum^{N}_{n=1}j_{n}\right)^{2}$, i.e., when $\left\vert\sum^{N}_{n=1}i_{n}\right\vert = \left\vert\sum^{N}_{n=1}j_{n}\right\vert$. Note that $\sum^{N}_{n=1}j_{n}$ are the eigenvalues of the total Pauli operator associated with the eigenstates $\vert j_{1},j_{2},\dots,j_{n}\rangle$.
Thus, if the initial state of the $N$-qubit system is a coherent superposition of eigenstates of the $\sigma^{\left(T\right)}_{z}$ operator, whose eigenvalues are equal in modulus, the condition $\left\vert\sum^{N}_{n=1}i_{n}\right\vert = \left\vert\sum^{N}_{n=1}j_{n}\right\vert$ is automatically satisfied and the fidelity dynamics does not oscillate at all.
On the other hand, a state of $N$ qubits that is not written in this way, i.e., a state that is written as a coherent superposition of the $\sigma^{\left(T\right)}_{z}$ eigenstates whose eigenvalues are not all equal in modulus (e.g., if exist at least one eigenvalue different from the others in modulus), has a fidelity which indeed oscillates in time. Consequently, in the quantum regime, the condition $\left\vert\sum^{N}_{n=1}i_{n}\right\vert \ne \left\vert\sum^{N}_{n=1}j_{n}\right\vert$ is a necessary condition for the non-monotonical behavior of the fidelity to take place. It is important to emphasize that this behavior is intrinsic to the geometry of the initial condition, that is, it depends only on the basis vectors spanning the initial state, and this property is not correlated with the initial entanglement. A simple example is the two-qubit state given by
\begin{equation}
\vert\phi\rangle = \vert1\rangle\otimes\left(\frac{\vert0\rangle + \vert1\rangle}{\sqrt{2}}\right) = \frac{1}{\sqrt{2}}\left(\vert10\rangle + \vert11\rangle\right). \label{FD.15}
\end{equation}
Although disentangled, the state is written as a coherent superposition of eigenstates of the $\sigma^{\left(T\right)}_{z}$ operator whose eigenvalues have different  moduli and, therefore, its fidelity oscillates in time following the equation below
\begin{equation}
F_{\vert\phi\rangle} = \frac{1}{2} +\frac{1}{2}e^{-4\Gamma\left(t,T\right)}\cos\left[4\Theta\left(t\right)\right]. \label{FD.16}
\end{equation}

\section{FIDELITY DYNAMICS IN AN MBQC}

From now on we will be concerned with the MBQC fidelity dynamics where the cluster states are subject to a dephasing channel. We will show how our previous result brings crucial consequences to the computational outcomes we can obtain, depending on the moment we decide to apply our set of projective measurements. To elucidate these aspects we analyze some common single qubit gates \cite{chuang} under the MBQC scheme \cite{briegel01, briegel02}. Following reference \cite{briegel01}, an arbitrary rotation can be achieved in a chain of five disentangled qubits,
\begin{equation}
|\Phi_{ini}\rangle = \left\vert\psi_{\textrm{in}}\right\rangle_{1}\otimes\vert +\rangle_{2}\otimes\vert +\rangle_{3}\otimes\vert +\rangle_{4}\otimes\vert +\rangle_{5}, \label{FD.17}
\end{equation}
where the qubits 2 to 5 are initially prepared in the state $\vert+\rangle_{n} = \frac{1}{\sqrt{2}}\left(\vert0\rangle_{n} + \vert1\rangle_{n}\right)$ while the qubit 1 is prepared in some input state which is to be rotated. We adopt the most general form  $\left\vert\psi_{\textrm{in}}\right\rangle_{1} = \alpha\left\vert0\right\rangle_{1} + \beta\left\vert1\right\rangle_{1}$ for the input state, where $\alpha$ and $\beta$ are complex numbers that satisfy the relation $\vert\alpha\vert^{2}+\vert\beta\vert^{2}=1$. To obtain the desired cluster state, the state (\ref{FD.17}) becomes entangled by an unitary operation $S$ (see references \cite{briegel01} and \cite{briegel03} for further details)
\begin{eqnarray}
S\vert\Phi_{ini}\rangle & = & \frac{1}{2} \left\vert\psi_{\textrm{in}}\right\rangle_{1}\vert 0\rangle_{2}\vert -\rangle_{3}\vert 0\rangle_{4}\vert-\rangle_{5} \nonumber \\
&-& \frac{1}{2} \left\vert\psi_{\textrm{in}}\right\rangle_{1}\vert 0\rangle_{2}\vert +\rangle_{3}\vert 1\rangle_{4}\vert +\rangle_{5} \nonumber \\
&-& \frac{1}{2} \left\vert\psi^{*}_{\textrm{in}}\right\rangle_{1}\vert 1\rangle_{2}\vert+\rangle_{3}\vert 0\rangle_{4}\vert-\rangle_{5} \nonumber \\
&+& \frac{1}{2} \left\vert\psi^{*}_{\textrm{in}}\right\rangle_{1}\vert 1\rangle_{2}\vert -\rangle_{3}\vert 1\rangle_{4}\vert +\rangle_{5}, \label{AP2.5}
\end{eqnarray}
where $\left\vert\psi^{*}_{\textrm{in}}\right\rangle_{1} = \sigma^{\left(1\right)}_{z}\left\vert\psi_{\textrm{in}}\right\rangle_{1} = \alpha\left\vert0\right\rangle_{1} - \beta\left\vert1\right\rangle_{1}$ and $\left\vert - \right\rangle_{n} = \frac{1}{\sqrt{2}}\left(\vert0\rangle_{n} - \vert1\rangle_{n}\right)$. The state $\left\vert\psi_{\textrm{in}}\right\rangle_{1}$ is rotated by measuring the qubits 1 to 4 one by one, and, at the same time that these measurements disentangle completely the cluster state (\ref{AP2.5}), they also implement a single qubit gate, ``printing" the outcome on qubit 5. An arbitrary rotation usually requires three angles in the Euler representation, the Euler angles, and it can be seen as a composition of three other rotations of the form: $U\left(\xi,\eta,\zeta\right)=U_{x}\left(\zeta\right)U_{z}\left(\eta\right)U_{x}\left(\xi\right)$, where $U_{k}\left(\phi\right)=\exp\left(\frac{-i\phi\sigma_{k}}{2}\right)$. In the MBQC scheme, the qubits $j=1,\ldots,4$ are measured in appropriately
chosen bases $\mathcal{B}_{j}\left(\phi_{j}\right) = \left\{\frac{\left\vert0\right\rangle_{j}+e^{i\phi_{j}}\left\vert1\right\rangle_{j}}{\sqrt{2}},\frac{\left\vert0\right\rangle_{j}-e^{i\phi_{j}}\left\vert1\right\rangle_{j}}{\sqrt{2}}\right\}$. Of course each measurement can possibly furnish two distinct results, ``up" (if the qubit $j$ is projected onto the first state of $\mathcal{B}_{j}\left(\phi_{j}\right)$) or ``down" (if the qubit $j$ is projected onto the other state of $\mathcal{B}_{j}\left(\phi_{j}\right)$), and the choice of the basis to measure the subsequent qubit depends on the previous results. In our examples we suppose (without incurring the risk of weakening our scheme) that all measurements give the result \textit{up}, which is not an event too rare, with the probability $1/16$ of occurrence. In this case, the first projector acting on the first qubit will be necessarily $\Pi_{1} = \vert+\rangle_{1}\langle+\vert$, irrespectively of the single qubit gate that we want to execute. However, the other three projectors $\Pi_{2}$, $\Pi_{3}$ and $\Pi_{4}$ that act on qubits 2, 3 and 4 are dependent on the specific choice of the single qubit gate, and they can be specified by the three Euler angles in terms of the first state of each one of the bases $\mathcal{B}_{2}\left(-\xi\right)$, $\mathcal{B}_{3}\left(\eta\right)$ and $\mathcal{B}_{4}\left(\zeta\right)$, respectively.

\begin{figure}[h]
\includegraphics[scale=0.30]{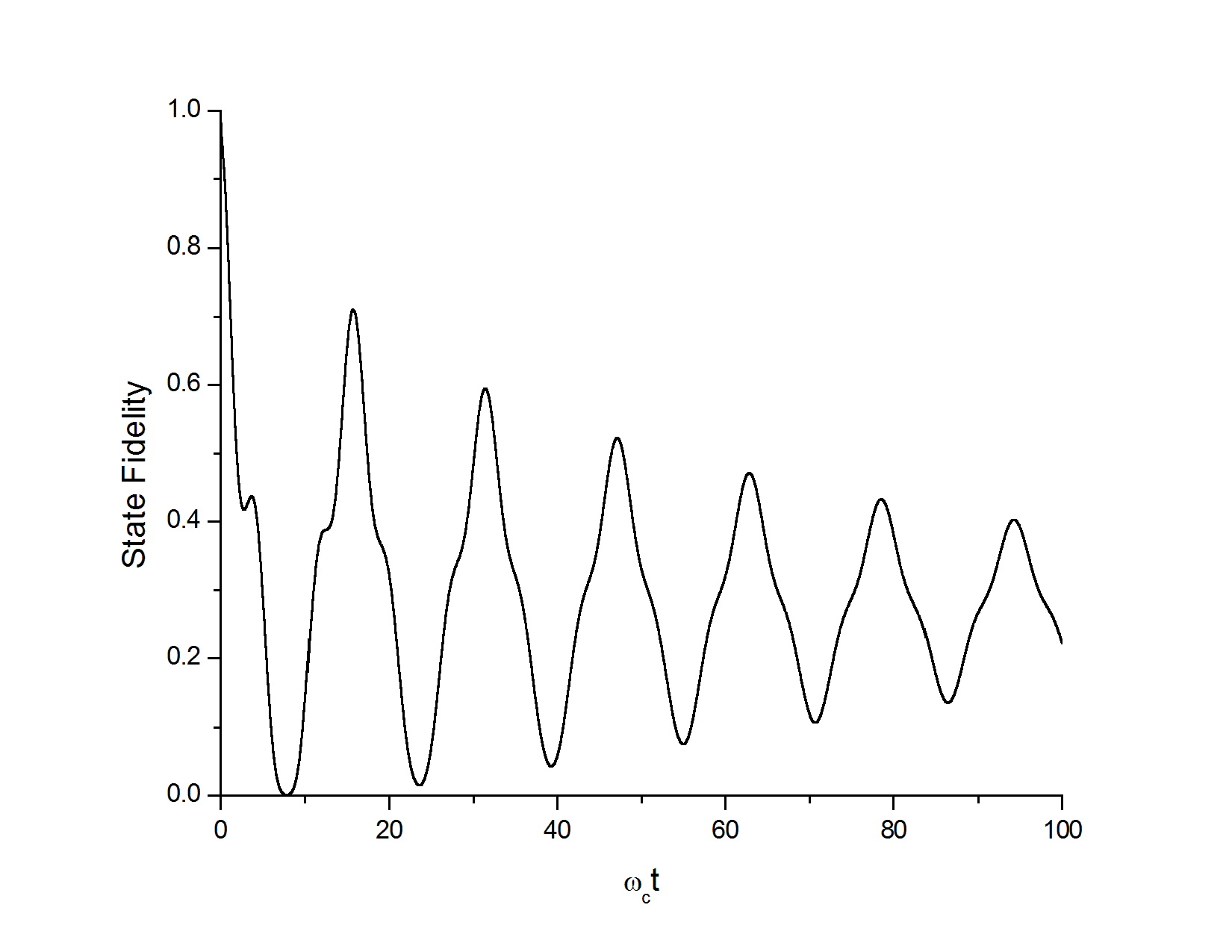}
\caption{ The state fidelity dynamics (\ref{QWQC.3}) for $\alpha=1$ in the quantum regime with $\Theta\left(t\right)$ and $\Gamma\left(t,T\right)$ given by equations (\ref{EDD.17C}) and (\ref{EDD.17D}) respectively. As we can see the oscillations is a characteristic feature in the quantum regime when the state is written as a coherent superposition of eigenstates of $\sigma^{\left(T\right)}_{z}$, whose eigenvalues are different in modulus. Here we consider $\eta=1/1000$, $\omega_{c}=100$, and $\omega_{T}=1$.}\label{statefidelity}
\end{figure}

After the first measurement, the resulting four qubit state that evolves under the influence of the environment is given by $\vert\psi\rangle_{2,...,5}(0)=\Pi_1 S\vert\Phi_{ini}\rangle$, where
\begin{eqnarray}
\vert\psi\rangle_{2,...,5}(0) & = & \frac{\alpha+\beta}{2\sqrt{2}}\vert 0\rangle_{2}\vert -\rangle_{3}\vert 0\rangle_{4}\vert -\rangle_{5} \nonumber \\
&-& \frac{\alpha+\beta}{2\sqrt{2}}\vert 0\rangle_{2}\vert +\rangle_{3}\vert 1\rangle_{4}\vert +\rangle_{5} \nonumber \\
&-& \frac{\alpha-\beta}{2\sqrt{2}}\vert 1\rangle_{2}\vert +\rangle_{3}\vert 0\rangle_{4}\vert -\rangle_{5} \nonumber \\
&+& \frac{\alpha-\beta}{2\sqrt{2}}\vert 1\rangle_{2}\vert -\rangle_{3}\vert 1\rangle_{4}\vert +\rangle_{5}. \nonumber \\
&& \label{OWQC.2}
\end{eqnarray}
Since the state (\ref{OWQC.2}) is written as a combination of eigenstates of the $\sigma^{\left(T\right)}_{z}$ with eigenvalues which are different in modulus, the fidelity dynamics associated with this state is an oscillatory function of time and is written as
\begin{widetext}
\begin{eqnarray}
F_{\vert\psi\rangle_{2,...,5}}\left(t\right) &=& \frac{3}{32}{e^{-16\Gamma\left(t,T\right)}}\cos\left[16\Theta\left(t\right)\right]+\frac{3}{8}{e^{-4\Gamma\left(t,T\right)}}\cos\left(4\Theta\left(t\right)\right) +\left[\frac{1}{16}-\frac{1}{32}\left(\alpha^{*}\beta+\alpha\beta^{*}\right)^{2}\right]{e^{-36\Gamma\left(t,T\right)}}\cos \left(12\Theta\left(t\right)\right)\nonumber\\
&+&\left[\frac{1}{16}+\frac{1}{32}\left(\alpha^{*}\beta+\alpha\beta^{*}\right)^{2}\right]{e^{-4\Gamma\left(t,T\right)}}\cos\left(12\Theta\left(t\right)\right) +\left[\frac{1}{128}-\frac{1}{128}\left(\alpha^{*}\beta+\alpha\beta^{*}\right)^{2}\right]{e^{-64\Gamma\left(t,T\right)}}\nonumber\\
&+&\left[\frac{1}{8}-\frac{1}{32}\left(\alpha^{*}\beta + \alpha\beta^{*}\right)^{2}\right]{e^{-16\Gamma{\left(t,T\right)}}} + \frac{5}{128}\left(\alpha^{*}\beta+\alpha\beta^{*}\right)^{2}+\frac{35}{128}.\nonumber\\ \label{QWQC.3}
\end{eqnarray}
\end{widetext}
The state fidelity (\ref{QWQC.3}) will present an oscillatory behavior in the quantum regime ($\omega_{c}\gg\omega_{T}$) where the exponential decay factor (\ref{EDD.17D}) plays a minor role compared with the oscillatory factor (\ref{EDD.17C}) and the quantum vacuum fluctuations contributes predominantly. Although Eq. (\ref{QWQC.3}) is written in terms of an input state that is a function of complex coefficients, there is no loss of generality if we regard them as real numbers. In Fig. \ref{statefidelity} we show the state fidelity dynamics (\ref{QWQC.3}) in the quantum regime assuming $\alpha=1$ and $\beta=0$, i.e., we choose the state $\left\vert\psi_{\textrm{in}}\right\rangle_{1}=\left\vert0\right\rangle_{1}$ as our input state. In the quantum regime for $\alpha=1$, Fig. \ref{statefidelity} shows that the state fidelity oscillates between maximum values, such as $71\%$ and $60\%$, at times $t=15,7/\omega_{c}$ and $t=31,4/\omega_{c}$ (where we have peaks)  and minimum values, such as $0,1\%$ and $1,5\%$, at times like $t=7,8/\omega_{c}$ and $t=23,5/\omega_{c}$ (where we have valleys).
But this oscillatory behavior is not a privilege of the input states $\left\vert\psi_{\textrm{in}}\right\rangle=\left\vert0\right\rangle_{1}$ or $\left\vert\psi_{\textrm{in}}\right\rangle=\left\vert1\right\rangle_{1}$, and more general input states will also present a qualitatively similar oscillatory behavior of the state fidelity dynamics, as we can see directly from Eq. (\ref{QWQC.3}). With this in mind, what can we say about the fidelity of quantum computation in this peculiar regime?

\begin{figure} [h]
\includegraphics[scale=0.45]{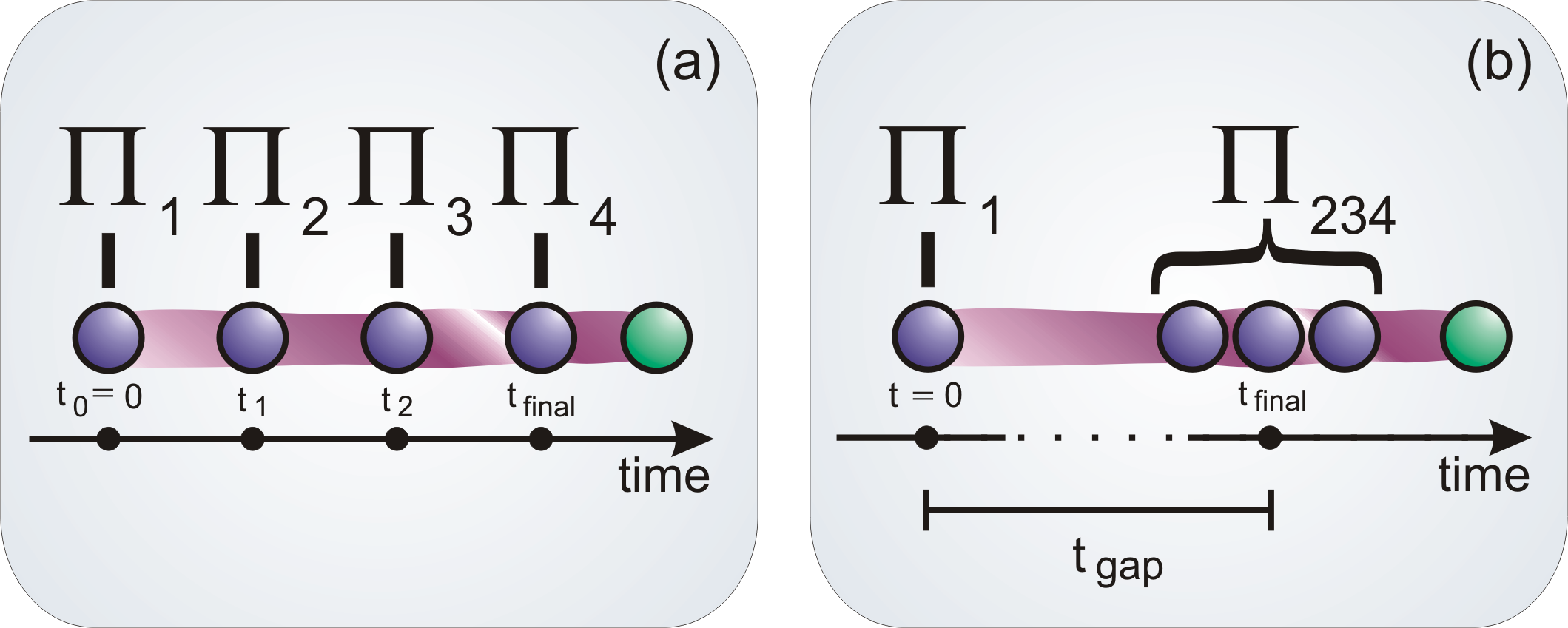}
\caption{Illustrative scheme of our considered scenarios. At $t_{0}=0$ we consider that the five-qubit state is already entangled and each qubit is ready to be measured. Besides, the first qubit is also projected at $t_{0}=0$. In (a) we suppose that the three subsequent measurements are applied at different instants of time and the qubits evolve non-unitarilly between the measurements. In (b), after wait a time gap, the other three subsequent measurements are made instantaneously at $t=t_{final}$. The result of the computation is ``printed" in the fifth qubit state (in green).}\label{Illustrative_scheme}
\end{figure}

The question above is relevant in the sense that in any realistic experimental realization, the construction of the state (\ref{FD.17}), the unitary operation of entanglement $S$, as well as the four subsequent projective measurements, are made in a finite time interval rather than instantaneously. Hence, it is worth  analyzing the computation when the system is subject to the deleterious effects caused by the environment. Here, we assume for simplicity that our initial state is given by (\ref{OWQC.2}), that is, our state at $t_{0}=0$ is the Cluster state (\ref{AP2.5}) prepared to perform the one-way quantum computation with the first measurement $\Pi_{1}$ already applied on qubit $1$. The subsequent measurements $\Pi_{2}$, $\Pi_{3}$ and $\Pi_{4}$ are supposed to be applied on qubits $2$, $3$, and $4$, respectively, in two different scenarios: in one of them the subsequent measurements are applied in sequence and at different instants of time, i.e., at time $t_{1}>t_{0}$ we apply the measurement $\Pi_{2}$, at time $t_{2}>t_{1}$ we apply the measurement $\Pi_{3}$ and finally at time $t_{final}>t_{2}$ we apply the measurement $\Pi_{4}$ (between the times $t_{1}$, $t_{2}$ and $t_{final}$ the system evolves according  to Eq. (\ref{EDD.15})); in the other scenario the measurements are applied in sequence ($\Pi_{2,3,4}=\Pi_{4}\Pi_{3}\Pi_{2}$) but practically at the same time -- a time that we call $t_{gap}$ (see Fig. \ref{Illustrative_scheme}).

With these two different scenarios in mind we are able to show our main result. Analyzing the implications of the state fidelity oscillations for the MBQC, we could verify that delayed measurements can in fact give better computational fidelity outcomes. As a matter of fact, there are time slots where we obtain better or worse computational fidelities defining periodic optimum waiting times. To clarify this assumption we analyze three different one qubit gate fidelities: the NOT gate, the HADAMARD gate and the PHASE gate.

\subsection{Measurements performed at different times}

Primarily, let us consider the first scenario where the measurements that characterize the specific one qubit gate are performed at different times and the state evolves non-unitarilly between measurements.

To begin with, we consider a NOT gate acting on an input state given by $\left\vert\psi_{\textrm{in}}\right\rangle_{1} = \left\vert0\right\rangle_{1}$. For this particular example the other three projectors are given by $\Pi_{2} = \vert-\rangle_{2}\langle-\vert$, $\Pi_{3} = \vert+\rangle_{3}\langle+\vert$ and $\Pi_{4} = \vert+\rangle_{4}\langle+\vert$, and the result of these projections would be represented by the output state $\left\vert\psi_{\textrm{out}}\right\rangle_{5} = \left\vert1\right\rangle_{5}$ if all measurements had been made before the environment starts its deleterious effect. Suppose, on the other hand, that these measurements are performed at later instants of time. Let us assume, for instance, that the projections are performed around the first valley of the state fidelity (see Fig. \ref{statefidelity}), within intervals  $\Delta t=2/\omega_{c}$ starting at $t = 6/\omega_{c}$, that is, $t_{1} = 6/\omega_{c}$, $t_{2} = 8/\omega_{c}$ and $t_{final} = 10/\omega_{c}$. Obviously we will not obtain a good fidelity for this computation since the state fidelity is very small within this time interval and, in fact, the NOT gate fidelity is $35,4\%$ showing that the probability of the output state be the desired state is approximately $35/100$. Now, if we consider that  $\Pi_{2}$, $\Pi_{3}$ and $\Pi_{4}$ are performed at $t_{1} = 14/\omega_{c}$, $t_{2} = 16/\omega_{c}$ and $t_{final} = 18/\omega_{c}$, i.e., are performed around the first peak of the state fidelity, the NOT gate fidelity is $53\%$. But we can obtain better results than these simply choosing another set of instants of time. If, for example, we set the controls to perform our measurements at slightly different times, choosing to apply the measurements in a smaller neighborhood around the first peak, we can get better results such as $84\%$ or $90\%$ for the set $\left(t_{1};t_{2};t_{final}\right)$ respectively at $\left(15,2/\omega_{c};15,7/\omega_{c};16,2/\omega_{c}\right)$ or $\left(15,5/\omega_{c};15,7/\omega_{c};15,9/\omega_{c}\right)$. Therefore, if we perform the measurements around the first valley we obtain a gate fidelity of $35,4\%$ whereas if we perform them at a later time, waiting to reach the surroundings of the fist peak, we obtain a much better gate fidelity. Another possibility can be imagined if we consider that the set of measurements is performed separately at each of the first three consecutive minima of the state fidelity, that is, we apply the first projector at the first valley, the second projector at the second valley and the third projector at the third valley of the state fidelity; in this case $\left(t_{1};t_{2};t_{final}\right) \approx \left(7,8/\omega_{c};23,4/\omega_{c};39/\omega_{c}\right)$ and the NOT gate fidelity assumes the value of $50\%$. On the other hand, if we do exactly the contrary, choosing the instants of time of the first three consecutive peaks, we obtain a NOT gate fidelity of $75,6\%$ at $\left(t_{1};t_{2};t_{final}\right) \approx \left(15,7/\omega_{c};31,4/\omega_{c};47,1/\omega_{c}\right)$.

Considering another example of one qubit gate acting on the same input state $\left\vert\psi_{\textrm{in}}\right\rangle_{1} = \left\vert0\right\rangle_{1}$, we can analyze the effect of the cluster state's oscillatory behavior on the MBQC fidelity in another situation of interest. As is well known, the HADAMARD gate transforms the state $\left\vert0\right\rangle$ into the state $\frac{1}{\sqrt{2}}\left(\left\vert0\right\rangle + \left\vert1\right\rangle\right)$, so, in an idealized situation, we would expect the output state $\left\vert\psi_{\textrm{out}}\right\rangle_{5}$ to assume the desired outcome $\left\vert\psi_{\textrm{out}}\right\rangle_{5}=\frac{1}{\sqrt{2}}\left(\left\vert0\right\rangle_{5} + \left\vert1\right\rangle_{5}\right)$. On the other hand, since our cluster state is interacting with the dephasing channel, the outcome of the MBQC can be very different from the expected one, mainly if we choose the wrong instants of time to apply the projective measurements, as we will see. For this particular example, we have $\Pi_{2} = \vert-,y\rangle_{2}\langle-,y\vert$, $\Pi_{3} = \vert+,y\rangle_{3}\langle+,y\vert$, and $\Pi_{4} = \vert+\rangle_{4}\langle+\vert$, where $\vert\pm,y\rangle = \frac{1}{\sqrt{2}}\left(\vert0\rangle\pm i\vert1\rangle\right)$. Admitting that these measurements are performed around the first valley of the state fidelity within intervals  $\Delta t=2/\omega_{c}$ starting at $t_{1} = 6/\omega_{c}$, as before, the HADAMARD gate fidelity is $39\%$, while if we consider that $t_{1} = 14/\omega_{c}$, $t_{2} = 16/\omega_{c}$ and $t_{final} = 18/\omega_{c}$, and the measurements are performed around the first peak, we have a HADAMARD gate fidelity of $52\%$. Nevertheless, a much better result can be obtained if $\left(t_{1},t_{2},t_{final}\right)\approx\left(15,5/\omega_{c},15,7/\omega_{c},15,9/\omega_{c}\right)$; in which case, the probability of the output state be the desired one is $85\%$. Now, consider again that the set of measurements is performed at the first three consecutive minima of the state fidelity; again $\left(t_{1};t_{2};t_{final}\right) \approx \left(7,8/\omega_{c};23,4/\omega_{c};39/\omega_{c}\right)$ and the gate fidelity assumes the value $50\%$. On the other hand, choosing the times of the first three consecutive peaks we get a HADAMARD gate fidelity of $71\%$ at $\left(t_{1};t_{2};t_{final}\right) \approx \left(15,7/\omega_{c};31,4/\omega_{c};47,1/\omega_{c}\right)$.

Finally, we examine another one qubit gate example that is often found in the literature. The PHASE gate under the MBQC can be accomplished with $\Pi_{2} = \vert+\rangle_{2}\langle+\vert$, $\Pi_{3} = \vert+,y\rangle_{2}\langle+,y\vert$, and $\Pi_{4} = \vert+\rangle_{4}\langle+\vert$, and the input state $\left\vert\psi_{\textrm{in}}\right\rangle_{1}=\frac{1}{\sqrt{2}}\left(\left\vert0\right\rangle_{1} + \left\vert1\right\rangle_{1}\right)$ is rotated, teleporting and printing the outcome in qubit number 5 whose output state $\left\vert\psi_{\textrm{out}}\right\rangle_{5}$ acquires a relative phase $i$ assuming the idealized value $\frac{1}{\sqrt{2}}\left(\left\vert0\right\rangle_{5} + i\left\vert1\right\rangle_{5}\right)$. Once again, taking into account the interaction of our ``quantum computer" with the dephasing environment, and applying the projectors at $t_{1} = 6/\omega_{c}$, $t_{2} = 8/\omega_{c}$ and $t_{final} = 10/\omega_{c}$, we obtain a gate fidelity of $48\%$ while if we wait to apply the projectors at $t_{1} = 14/\omega_{c}$, $t_{2} = 16/\omega_{c}$ and $t_{final} = 18/\omega_{c}$ we get a gate fidelity of $65\%$. However, applying the measurements in a smaller neighborhood of the first peak we can get a gate fidelity of $95\%$ at $t_{1} = 15,5/\omega_{c}$, $t_{2} = 15,7/\omega_{c}$ and $t_{final} = 15,9/\omega_{c}$. Considering that the set of measurements is performed at the first three consecutive minima of the state fidelity, at $t_{1} = 7,8/\omega_{c}$, $t_{2} = 23,4/\omega_{c}$ and $t_{final} = 39/\omega_{c}$, the value of the gate fidelity is $46\%$, while if the set of measurements is applied on the first three consecutive maxima we get $85\%$ for the gate fidelity.

Therefore, depending on the times we choose to perform our set of projective measurements we will obtain better or worse computational fidelity results.

\subsection{Measurements performed at the same time}

\begin{figure}[h]
\includegraphics[scale=0.30]{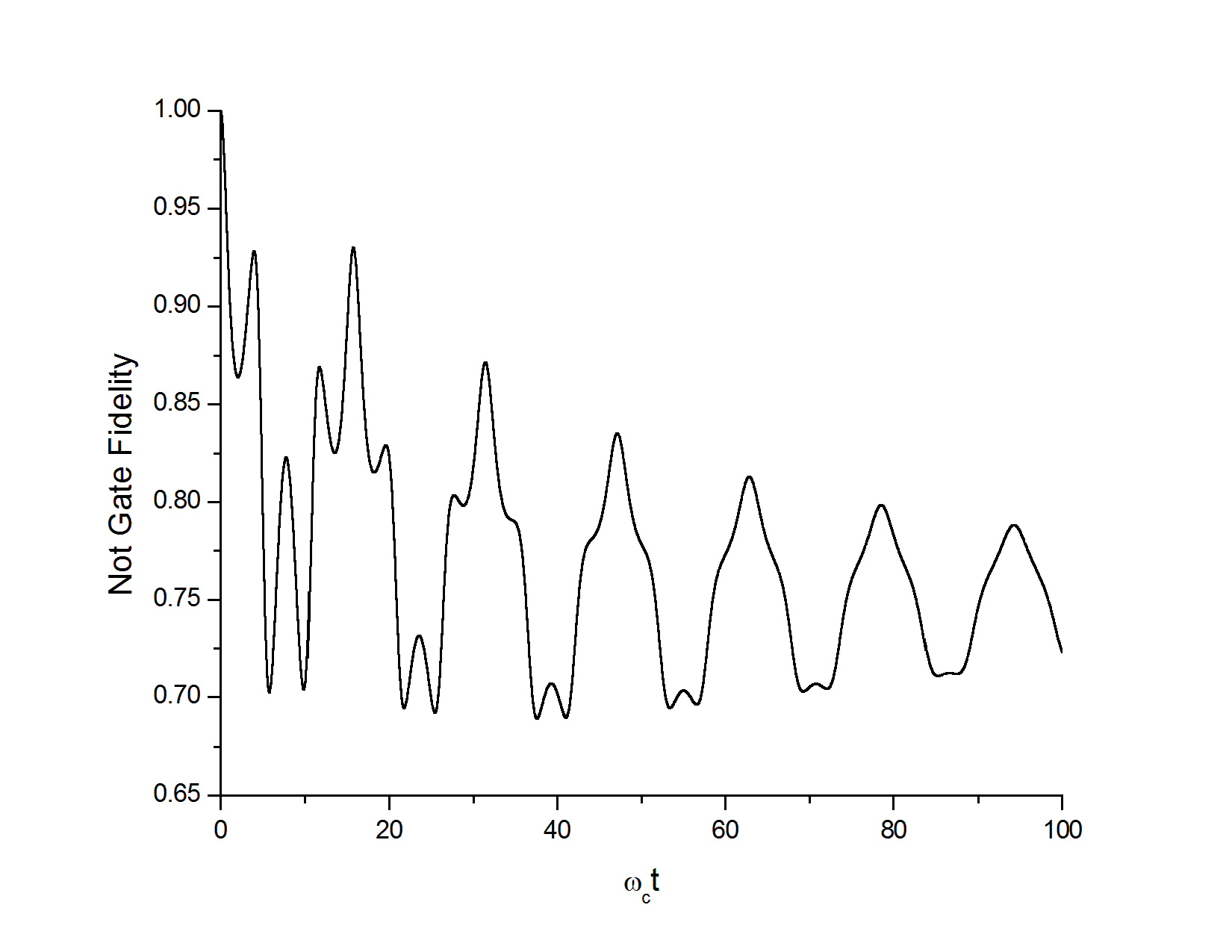}
\caption{Here we show the measurement-based quantum NOT gate fidelity as a function of $t_{gap}$ when the three subsequent measurements are performed almost simultaneously. We consider $\eta=1/1000$, $\omega_{c}=100$, and $\omega_{T}=1$.}\label{Fig1}
\end{figure}

Now, let us consider the other scenario where the subsequent measurements are performed in sequence but practically at the same time. Again we consider the same three examples of gate fidelities. We start with the NOT gate fidelity dynamics and in Fig. \ref{Fig1} we show this gate fidelity for $\alpha=1$ and $\beta = 0$ as a function of time in the quantum regime. As in our previous examples, we observe that there are times that maximize the value of the gate fidelity and times that minimize it. It is clear that if all measurements are performed at $t=0$ the computation fidelity is $100\%$, but if there is a gap between the first and the three subsequent measurements, then there is an optimum value of the time gap. In the example illustrated in Fig. \ref{Fig1}, if $t_{gap}$ is greater than $0,8/\omega_c$ (where the gate fidelity is $93\%$) the best gate fidelity is obtained for $t_{gap}=15,7/\omega_c$, when it reaches $93\%$ again. If we apply the same gate operation at later times like $t_{gap}=31,4/\omega_c$ or $t_{gap}=47,1/\omega_c$, we still get a gate fidelity better than $80\%$. However, if we apply this operation at $t_{gap}=5,8/\omega_c$ we obtain a gate fidelity of $70\%$, showing that fast measurements is not a warranty of good MBQC results.

\begin{figure}[h]
\includegraphics[scale=0.30]{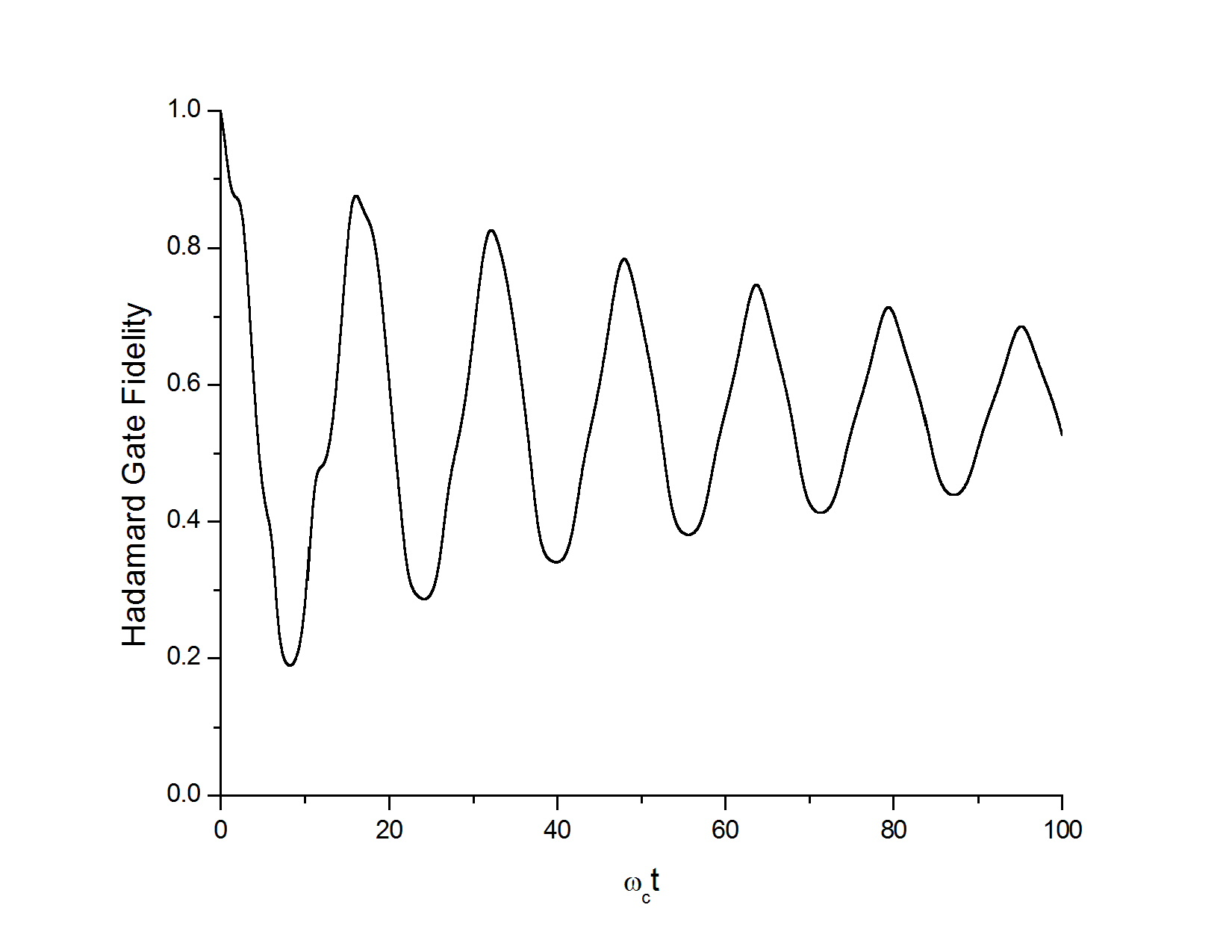}
\caption{Here we show the measurement-based quantum HADAMARD gate fidelity as a function of $t_{gap}$ when the three subsequent measurements are performed almost simultaneously. We choose the state $\vert\psi_{\textrm{in}}\rangle_{1}=\vert0\rangle_{1}$ to be rotated to $\vert\psi_{\textrm{out}}\rangle_{5}=\frac{1}{\sqrt{2}}\left(\vert0\rangle_{5}+\vert1\rangle_{5}\right)$. We consider $\eta=1/1000$, $\omega_{c}=100$, and $\omega_{T}=1$.}\label{Fig2}
\end{figure}

Considering a HADAMARD gate fidelity dynamics, we can see from Fig. \ref{Fig2} that a input state $\left\vert\psi_{\textrm{in}}\right\rangle_{1} = \left\vert0\right\rangle_{1}$ is rotated to the output state $\left\vert\psi_{\textrm{out}}\right\rangle_{5} = \frac{1}{\sqrt{2}}\left(\left\vert0\right\rangle_{5} + \left\vert1\right\rangle_{5}\right)$ with probability greater than $80\%$ at times like $t_{gap}=15,7/\omega_c$, $t_{gap}=31,4/\omega_c$ or $t_{gap}=47,1/\omega_c$ while it is rotated to the same output state with probability of less than $40\%$ at times like $t_{gap}=7,8/\omega_c$, $t_{gap}=23,5/\omega_c$ or $t_{gap}=39,2/\omega_c$.
\begin{figure} [h]
\includegraphics[scale=0.30]{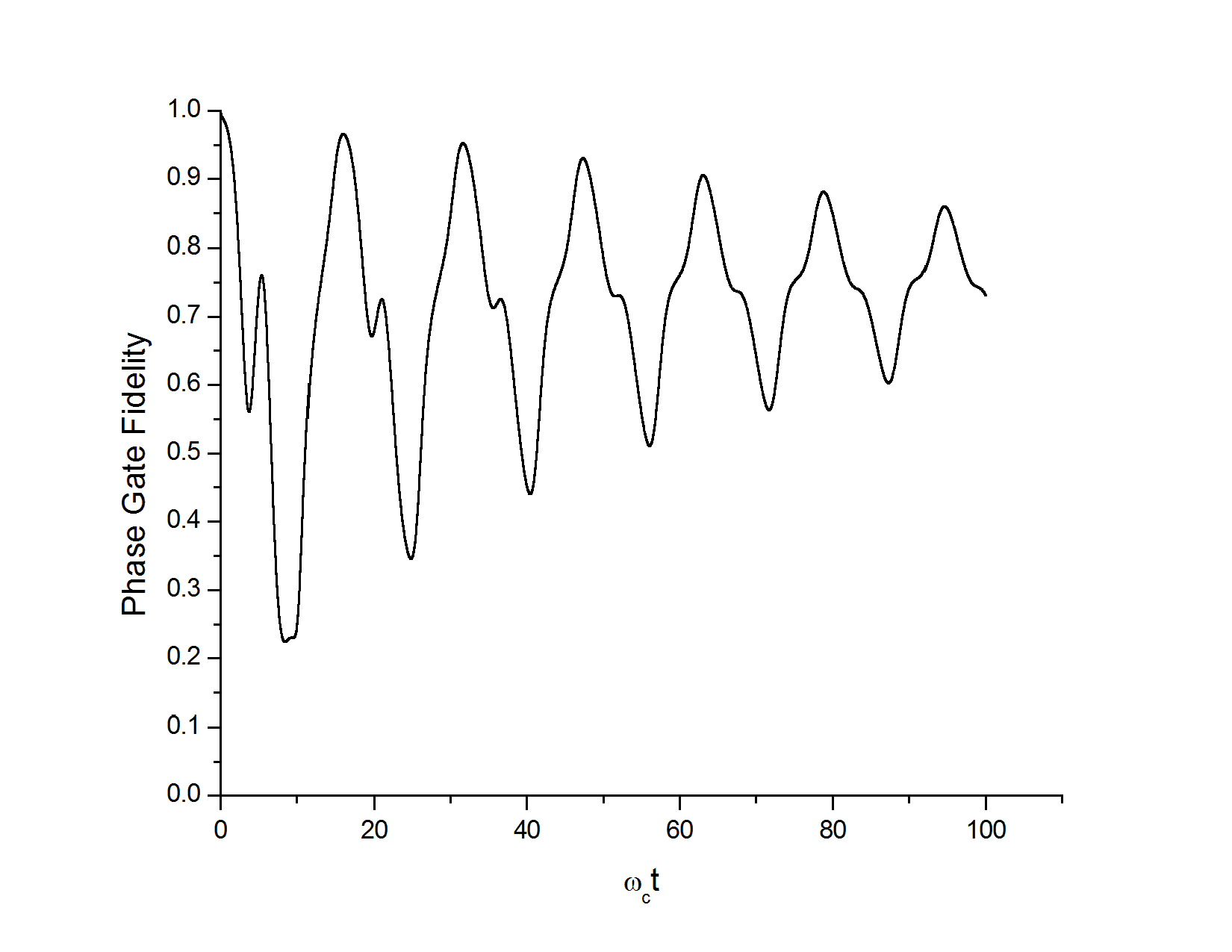}
\caption{Here we show the PHASE gate fidelity computed under the one-way quantum computation scheme as a function of $t_{gap}$ for the quantum regime. We choose the state $\vert\psi_{\textrm{in}}\rangle_{1}=\frac{1}{\sqrt{2}}\left(\vert0\rangle_{1}+\vert1\rangle_{1}\right)$ to be rotated to $\vert\psi_{\textrm{out}}\rangle_{5}=\frac{1}{\sqrt{2}}\left(\vert0\rangle_{5}+i\vert1\rangle_{5}\right)$. We consider $\eta=1/1000$, $\omega_{c}=100$, and $\omega_{T}=1$. }\label{Fig3}
\end{figure}

In Fig. \ref{Fig3} we can see the PHASE gate fidelity dynamics and observe once again that the oscillatory behavior presented by the cluster state interacting with the specific kind of quantum channel considered in this paper produces instants of time that optimize the value of the computation compared with times that do exact the opposite. Again we see that at times like $t_{gap}=15,9/\omega_c$, $t_{gap}=31,6/\omega_c$ or $t_{gap}=47,3/\omega_c$ we have a gate fidelity of $96\%$, $95\%$ and $93\%$, respectively, while at times such as $t_{gap}=8,4/\omega_c$, $t_{gap}=24,8/\omega_c$ or $t_{gap}=40,4/\omega_c$ we have a gate fidelity of  $22\%$, $34\%$ and $44\%$.

It is important to emphasize that ultra fast measurements, which have to be performed in the very short bath correlation time scale, can be produced with current technology \cite{expDD, expUDD}. These are the basis of the dynamical decoupling techniques that are applied to beat the decoherence process \cite{DD}. Furthermore, even in this very short time scale, the time that each measurement is applied can be very precise, as we can see, for example, in the experimental realization of the Uhrig dynamical decoupling, the Carr-Purcell-Meiboom-Gill-style multi-pulse spin echo \cite{expUDD}, and others. This implies that the scenario studied in the manuscript is very realistic and that any MBQC realized with ultra fast measurements needs to account for the oscillatory behavior of the dynamics.

\section{CONCLUSION}

We study the exact dynamics of an $N$-qubit system interacting with a common dephasing environment and we introduce a necessary condition for the system fidelity to present a non-monotonical behavior. Our approach reveals that this characteristic does not depend on the initial quantum entanglement and, in fact, is a property connected with the geometry of the state. Actually, for any initial state given by a superposition of eigenstates of the total Pauli $\sigma^{\left(T\right)}_{z}$ operator, the fidelity exhibits a non-monotonical character if at least one of the eigenvalues of the components differs from the others. We show that this behavior of the fidelity brings crucial implications to the MBQC, that is, we show that, under the action of a common dephasing environment, this non-monotonical time dependence can provide us with appropriate time intervals for the preservation of better computational fidelities. We have illustrated our findings by examining the fidelity of a NOT, a HADAMARD and a PHASE quantum gates realized via MBQC.

\end{document}